# Johari-Goldstein relaxation far below $T_g$:
# Experimental evidence for the Gardner transition in structural glasses?

K. Geirhos, P. Lunkenheimer*, and A. Loidl

*Experimental Physics V, Center for Electronic Correlations and Magnetism, University of Augsburg, 86159 Augsburg, Germany*

Experimental evidence for the Gardner transition, theoretically predicted to arise deep in the glassy state of matter, is scarce. At this transition, the energy landscape sensed by the particles forming the glass is expected to become more complex. In the present work, we report the dielectric response of two typical glass formers with well-pronounced Johari-Goldstein $\beta$ relaxation following this response down to unprecedented low temperatures, far below the glass transition. As the Johari-Goldstein process is believed to arise from the local structure of the energy landscape, its investigation seems an ideal tool to seek evidence for the Gardner transition. Indeed, we find an unusual broadening of the $\beta$ relaxation below $T_G \approx 110$ K for sorbitol and $T_G \approx 100$ K for xylitol, in excess of the expected broadening arising from a distribution of energy barriers. Thus, these results provide hints at the presence of the Gardner transition in canonical structural glass formers.

When a liquid is cooled sufficiently fast ("supercooled"), crystallization can be avoided and a glass is formed with an amorphous structure. This occurs at the glass temperature $T_g$, where the increasing viscosity becomes so high that the material can be regarded as solid. However, in contrast to the crystallization temperature, the glass temperature is only a kinetic phenomenon and does not mark a phase transition but merely the loss of thermodynamic equilibrium under cooling with reasonable cooling rates. Nevertheless, in recent years there is growing evidence that in fact there is an underlying phase transition that governs many of the unusual universal properties of glass-forming liquids and glasses (see, e.g., [1,2,3,4,5]) as predicted, e.g., by the random-first-order transition theory [6,7]. It is supposed to occur at a kind of "ideal" glass-transition temperature far below $T_g$ and may be identified by the Kauzmann temperature $T_K$, where the entropy of an infinitely slowly supercooled liquid would fall below that of the corresponding crystal [3,8], or by the Vogel-Fulcher temperature $T_{VF} \approx T_K$, where the extrapolated viscosity of the supercooled liquid would diverge [3,9,10]. Nevertheless, experimentally this ideal glass transition is not accessible even for the most patient experimenter as the material is falling out of thermodynamic equilibrium for any reasonable cooling rate.

Interestingly, very recently theoretical predictions of another phase transition supposed to occur deep in the glassy state, at a temperature $T_G$ below $T_g$ and even below $T_K$, have come into the focus of interest [11,12,13,14,15,16,17]. It is termed Gardner transition and was originally inferred from the mean-field theory for spin glasses [18,19]. In contrast to the phase transition close to $T_K$, leading to an ideal glass state, the Gardner transition is also expected to be observable in non-equilibrium [20], which is inevitable when cooling a liquid far below $T_g$ with reasonable rate. In several current works it was shown that the Gardner transition is not only expected in spin glasses [18,19] but also in hard-sphere and other model glasses for infinite [11,12,21] and three dimensions [13,14,22]. It was further argued [17] that the Gardner transition is not limited to glasses or jammed packings, but likely is of prime importance in a variety of disordered systems.

To understand the implications of this transition, the energy-landscape picture [23,24,25] can be invoked: Each basin in this landscape represents a separate amorphous glass state. The particle motions that govern viscous flow, the so-called $\alpha$ relaxation, correspond to jumps between these "metabasins". They become increasingly slow when temperature is lowered and, finally, the system gets stuck within one of the basins, i.e., it becomes a glass. For infinitely slow cooling rate, this would occur via the "ideal" glass transition, representing a phase transition into the lowest-lying glass state. Within this energy-landscape picture, the Gardner transition is assumed to give rise to sub-basins within the metabasins, which exhibit a fractal hierarchy, leading to a so-called marginal fractal glass state below $T_G$ [11,14].

Interestingly, sub-basins separated by small barriers within the metabasins are often also supposed to explain the occurrence of Johari-Goldstein (JG) $\beta$-relaxation processes. This type of secondary relaxation, which is faster than the $\alpha$ relaxation, is an intrinsic property of glass-forming matter [26,27,28] and assumed to arise from local molecular motions on a smaller length scale than the $\alpha$ relaxation [29,30]. However, in contrast to the Gardner scenario mentioned above, the sub-basins connected by the JG relaxation processes obviously already exist at elevated temperatures because this relaxation is commonly observed at temperatures below and around $T_g$, i.e., above $T_G$ [26,27,28,31,32].

The temperature-dependent development of the JG relaxation in dielectric loss spectra is schematically indicated in Figs. 1(a)-(d): Below a merging temperature (often found to be similar to the characteristic temperature $T_c$ of the mode-coupling theory [33]), it shows up as a well-separated second peak at a higher frequency than the dominating $\alpha$-relaxation peak [Fig. 1(b)]. Because of its weaker temperature dependence, which follows Arrhenius behavior instead of the usual super-Arrhenius behavior of the $\alpha$ relaxation [cf. the relaxation times $\tau$ shown in Fig. 1(e)], it becomes well separated from the dominating $\alpha$-relaxation peak when temperature decreases and finally dominates the spectra below $T_g$ [Fig. 1(c)]. The width of the $\beta$ peak is known to become successively broader when temperature is lowered.

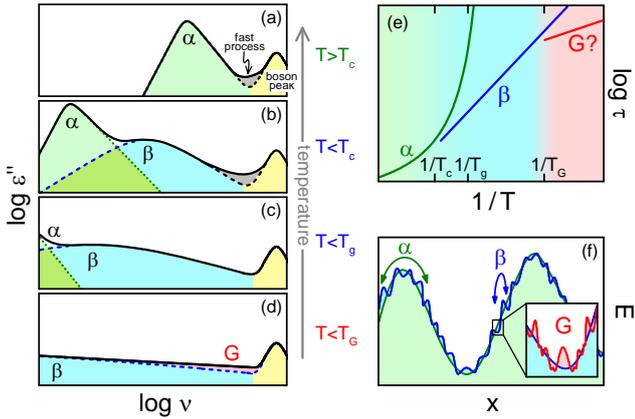

FIG. 1. (a-d) Schematic loss spectra of glass forming materials in different temperature regimes: (a) The liquid state above $T_c$. (b) The supercooled liquid, $T_g < T < T_c$. (c) The glass below $T_g$. (d) The glass far below $T_g$ and below the Gardner transition, $T < T_G < T_g$. The different contributing dynamic processes are indicated by color: The $\alpha$ relaxation (green), the JG $\beta$-relaxation (blue), the fast process (grey), and the boson peak (yellow) [35,37]. The red area in (d) indicates the suggested additional contribution arising from the sub-basins in the energy landscape induced by the Gardner transition. (e) Typical temperature dependence of the $\alpha$- and $\beta$-relaxation times in an Arrhenius plot with a possible additional Gardner relaxation arising below $T_G$. (f) Schematic view of the energy-landscape scenario assumed to explain the occurrence of the JG $\beta$-relaxation [24,25,42]. Inset: Possible modification of the local $\beta$-relaxation basins by the Gardner transition, leading to a fractal roughening of the landscape.

Until now, the relation of the Gardner transition to the JG relaxation has only been briefly discussed [11,13,14] and it is not clear how the breaking up of the metabasin into a fractal hierarchy of sub-basins arising at the Gardner transition is related to the fine structure of the energy landscape assumed to cause the JG $\beta$-relaxation. Clearly, the two features cannot be the same because, as mentioned above, the JG relaxation is commonly observed at elevated temperatures. If we assume that the Gardner transition leads to a further fractal "roughening" of the existing $\beta$-relaxation sub-basins [inset of Fig. 1(f)], an impact on the JG relaxation seems likely and even a separate "Gardner relaxation" may arise [Fig. 1(e)].

So far all reports on Gardner phenomena were either based on theoretical considerations or numerical simulations. To our knowledge, until now there is only a single experimental work providing evidence for the Gardner transition investigating a granular, two-dimensional glass [15]. It is an interesting question if the Gardner transition can also be detected in canonical structural glass-formers, e.g., the molecular glass-forming liquids whose investigation in the past decades has proven very useful in learning more about the glass transition [24,34,35,36,37]. Compared to the numerous investigations around and above $T_g$ of such glass formers, measurements down to temperatures far below $T_g$ are relatively rarely done. Nevertheless, it is clear that in the temperature dependence of quantities as the specific heat, volume, or the dielectric permittivity, so far no obvious anomalies were reported that would point to a phase transition below $T_g$ [38,39,40,41] (see also Fig. S1 of the Supplemental Material). Obviously, in these canonical glass formers the signature of the Gardner transition must be of much subtler nature and probably becomes smeared out, in a similar way as the mode-coupling $T_c$ becomes smeared out by thermally activated processes

In the present work, we tackle this problem by investigating the JG $\beta$-relaxation of glass-forming liquids down to unprecedented low temperatures by dielectric spectroscopy. As pointed out above, the JG $\beta$-relaxation is associated with jumps between local energy minima forming the fine structure within the energy landscape [Fig. 1(f)] [24,25,42]. Thus, this dynamic process arises from the local structure of the energy landscape within one metabasin. As just this landscape is predicted to change at the Gardner transition [inset of Fig. 1(f)], the investigation of the JG relaxation process, which senses the energy landscape, at temperatures far below $T_g$ could be an ideal tool to check for this transition.

Xylitol was purchased from Aldrich with a purity $\geq 98\%$ and sorbitol from AppliChem with a purity of 99.7% and used without further purification. To obtain high-resolution dielectric measurements down to He temperature, three techniques were combined [43]: At $10^{-2} \leq \nu \leq 10^7$ Hz, a frequency response analyzer (Novocontrol Alpha-A analyzer) was used. Data with highest precision at the lowest temperatures were collected with an Andeen-Hagerling AH2700A capacitance bridge ($50 \leq \nu \leq 2\times10^4$ Hz). For these two methods the samples were prepared as parallel-plate capacitors. Additional measurements up to higher frequencies ($10^3 \leq \nu \leq 10^8$ Hz) were performed with a coaxial reflectometric technique employing the impedance analyzer Agilent 4294A [44]. For sample cooling, a $^4$He-bath cryostat (Cryovac), a $N_2$-gas cryostat (Novocontrol), and a closed-cycle refrigerator (CTI-cryogenics) were used.

The investigated glass formers xylitol ($T_g = 248$ K) and sorbitol ($T_g = 274$ K) are known to exhibit well-pronounced JG $\beta$ relaxations [28,45,46], which were identified as "genuine" JG relaxations [45,46]. Broadband dielectric-loss spectra of these materials revealing the $\alpha$ relaxation at high temperatures and the successive evolution of the $\beta$ relaxation with decreasing temperature can be found, e.g., in Refs. [28,37,40,41,47,48]. However, to our knowledge measurements extending down to He temperatures of these or of other glass formers with pronounced $\beta$ relaxation have never been reported. Figures 1(a) and (b) show spectra of the dielectric constant $\varepsilon'$ and loss $\varepsilon''$, respectively, as measured for sorbitol in the frequency region dominated by the $\beta$ relaxation and at temperatures from 14 K below $T_g$ down to 5 K. Corresponding spectra for xylitol are provided in Fig. S2 of the Supplemental Material [49]. As can be deduced, e.g., from the investigations of Ref. [50] (including xylitol), even the highest temperatures considered here are sufficiently below $T_g$ to exclude any detectable time dependence from physical aging.

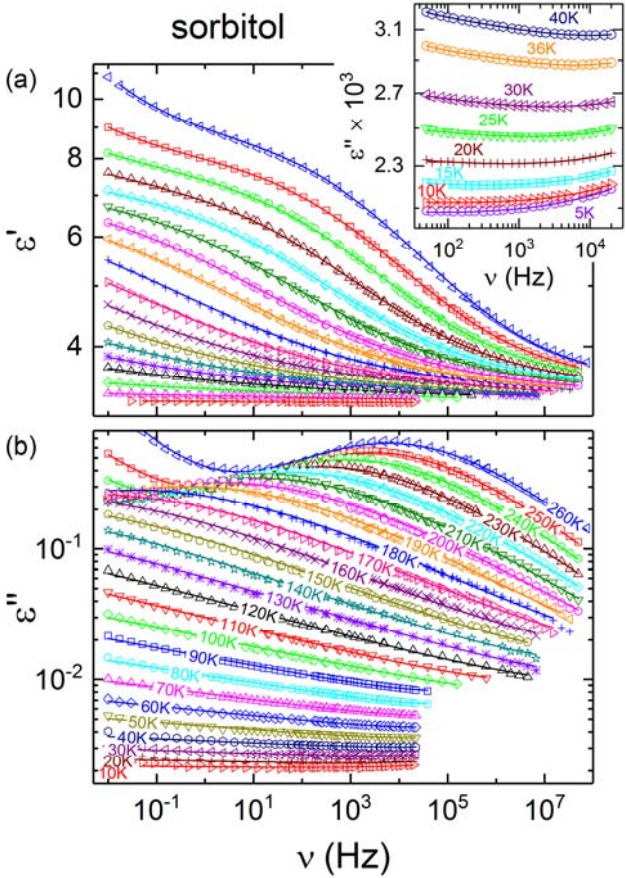

FIG. 2. Spectra of the dielectric constant (a) and loss (b) of sorbitol for various temperatures. The lines are fits by a combination of power laws and the CC function as described in the text. For clarity reasons, in (a) at low temperatures curves are shown for less temperatures than in (b). The inset shows a magnified view of the loss at the lowest temperature, in the region of the minimum.

In Fig. 2, at the highest temperatures, the typical steplike decrease in $\varepsilon'(\nu)$ and peak in $\varepsilon''(\nu)$ signify the well-known JG $\beta$ relaxation of this material. The additional increase showing up at low frequencies in both quantities corresponds to the high-frequency tail of the $\alpha$ relaxation, whose main spectral features lie outside of the shown frequency window [28] [cf. Fig. 1(c)]. Following thermally activated behavior below $T_g$ [cf. Figs. 1(e) and S3], the $\beta$ relaxation shifts to lower frequencies with decreasing temperature. In the loss spectra [Fig. 2(b)], below about 150 K only its high-frequency flank remains visible showing up as a power law $\nu^{1-\alpha}$ [cf. Fig. 1(d)]. This power law becomes exceedingly shallow at temperatures below about 100 K. Finally, $\varepsilon''$ starts to exhibit an additional increase with frequency and a minimum develops (inset of Fig. 2). At these low temperatures, $\varepsilon'(\nu)$ [Fig. 2(a)] is dominated by the high-frequency dielectric constant $\varepsilon_\infty$. For xylitol, very similar overall behavior of $\varepsilon'(\nu)$ and $\varepsilon''(\nu)$ is detected (see Supplemental Material, Fig. S2 [49]).

For $\beta$ relaxations, the empirical Cole-Cole (CC) function, $\varepsilon^* = \varepsilon' - i\varepsilon'' = \varepsilon_\infty + \Delta\varepsilon / [1+(\omega\tau)^{1-\alpha}]$ [51] is known to provide a good description of the dielectric spectra [28,48] ($\Delta\varepsilon$ is the relaxation strength and $\omega = 2\pi\nu$ the circular frequency). The introduction of the width parameter $\alpha$ (with $0 \leq \alpha < 1$) in the CC equation leads to a symmetric broadening of the spectral relaxation features compared to the Debye function, which is recovered for $\alpha = 0$. In $\varepsilon''(\nu)$ the CC function leads to power laws $\varepsilon'' \propto \nu^{1-\alpha}$ and $\nu^{\alpha-1}$ at the low and high-frequency flanks of the peak, respectively. The lines in Fig. 2 at $T \geq 210$ K are fits, simultaneously performed for $\varepsilon'$ and $\varepsilon''$, using the sum of a CC function and an additional power-law decrease, the latter accounting for the high-frequency tail of the $\alpha$ relaxation seen at low frequencies. When the $\beta$ peak has shifted out of the frequency window at lower temperatures ($T \leq 100$ K), it was sufficient to only use a power law $\varepsilon'' = c\nu^{\alpha-1}$ for the fits of its right flank. Accounting for $\varepsilon_\infty$ and considering the Kramers-Kronig relation then leads to $\varepsilon' = \varepsilon_\infty + c\,\nu^{\alpha-1} \tan[\alpha\pi/2]$ for the real part [52].

The mentioned increase of $\varepsilon''(\nu)$ with frequency, revealed at the lowest temperatures leading to a minimum (inset of Fig. 2), was taken into account by a power law $\nu^a$ with a positive exponent $a$ varying between 0.1 and 0.4. This leads to a $-\cot(a\pi/2)$ prefactor in $\varepsilon'(\nu)$. We found it necessary to include this positive power law up to temperatures of 190 K. Interestingly, such a power-law increase is predicted for thermally activated hopping over asymmetric double-well potentials [53], which are the same potentials known to lead to the universal tunneling-induced anomalies, typical for glasses at very low temperatures [54]. Indeed, this mechanism was invoked to explain the positive power law observed in dielectric-loss spectra around 1 kHz far below $T_g$ for several glass formers [39] (however, not including a material with well-pronounced $\beta$ relaxation). A similar explanation was suggested for a power law that was detected in the susceptibility deduced from light-scattering investigations of various glass formers at frequencies above 10 GHz and temperatures below $T_g$ [55,56]. Anomalies observed in dielectric measurements at very low temperatures [38,39] were in addition also ascribed to tunneling *through* these barriers and a power-law increase of $\varepsilon''(\nu)$ was also detected in this regime [39]. A detailed investigation of these effects is out of the scope of the present work and in the following we will concentrate on the development of the $\beta$ relaxation far below $T_g$.

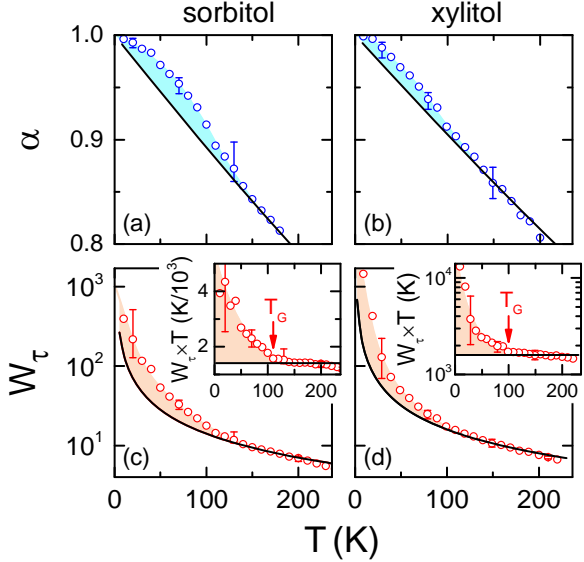

FIG. 3. Circles in (a) and (b): temperature dependence of the width parameter $\alpha$ as obtained from fits of the dielectric spectra of sorbitol and xylitol, respectively. Circles in (c) and (d): Corresponding half widths of the relaxation-time distribution. The insets show the same data multiplied by $T$. The lines in (c) and (d) (including the insets) are $W_\tau \propto 1/T$ laws with the proportionality factor adapted to match the experimental data at high temperatures. From these, the lines in (a) and (b) were calculated. The colored regions indicate the additional broadening ascribed to the roughening of the energy landscape arising at the Gardner transition.

The circles in Figs. 3(a) and (b) show the temperature dependence of the width parameter $\alpha$ as resulting from the fits of the dielectric spectra of sorbitol (Fig. 2) and xylitol (Fig. S2), respectively. With decreasing temperature, it continuously increases and finally reaches values close to one, which implies extremely shallow flanks of the $\beta$-relaxation peaks. The broadening of relaxation peaks (compared to the Debye function) is commonly ascribed to the heterogeneity of amorphous matter, leading to a distribution of relaxation times $g_\tau(\tau)$ [57,58,59]. From the width parameter $\alpha$ of the CC function, the half with $W_\tau$ (in decades) of this distribution can be calculated [51,60], which is shown for the two glass formers in Figs. 3(c) and (d). The extremely broad $\beta$-relaxation peaks, deduced from the observed very shallow power laws, also imply extremely broad distribution functions reaching half widths up to hundreds (sorbitol) or even more than 1000 decades (xylitol) at the lowest temperatures. At first glance, this seems quite unrealistic and one may be tempted to invoke additional spectral contributions as the nearly constant loss, which is considered, e.g., within the extended coupling model [61]. However, one should be aware that the originally purely empirical CC function can be well approximated when assuming that the distribution of relaxation times $g_\tau(\tau)$ itself arises from a Gaussian distribution $g_E(E)$ of the energy barriers governing the molecular motion [60]. In this case, it is known that a temperature *independent* $g_E$ leads to a strongly temperature *dependent* $g_\tau$ distribution whose half width $W_\tau$ diverges with $1/T$ [27,60] (see discussion of relaxation-time map in the Supplementary Information [49] for a rationalization of this notion). Especially, its temperature dependence is given by $W_\tau = A\sigma/(k_B T)$, where $A \approx 1.023$ [60] and $\sigma$ is the temperature-independent standard deviation of $g_E$. Thus, the extreme broadening of the $\beta$ relaxation suggested by the lowest-temperature loss spectra is naturally explained in this way. The expected $W_\tau \propto 1/T$ behavior is indicated by the solid lines in Figs. 3(c) and (d), which indeed well reflect the general trend of the experimental data. It should be noted that only a single parameter (the standard deviation $\sigma$) is available to adapt these curves to the experimental results. For both materials, $\sigma$ is found to be about 20% of the activation energies $E = 0.62$ eV (sorbitol) and $E = 0.60$ eV (xylitol) of the JG $\beta$ process. Remarkably, while the high-temperature data can be well matched in this way, below about 110 K (sorbitol) or 100 K (xylitol), the width starts to increase stronger than expected. This becomes especially obvious in the insets of Fig. 3(c) and (d) showing $W_\tau$ multiplied by $T$. For the width parameter $\alpha(T)$, the $1/T$ divergence of $W_\tau$ corresponds to approximately linear behavior as shown by the solid lines in Figs. 3(a) and (b). Here corresponding deviations of the experimental data show up, namely, at low temperatures $\alpha$ is found to be larger than predicted.

Overall, while the general trend of an extreme broadening of the $\beta$ relaxation in sorbitol and xylitol at low temperatures can be well understood assuming a Gaussian distribution of energy barriers, at low temperatures there seems to be an additional contribution to the observed broadening. Within the Gardner-transition scenario discussed in section I, this finding can be explained in two ways: (i) The energy landscape just becomes more complex below the Gardner transition, leading to an additional contribution to the broadening of the relaxation-time distribution and, thus, to an additional broadening of the $\varepsilon''(\nu)$ curves [red area in Fig 1(d)]. (ii) Just as local minima are generating the $\beta$ relaxation [Fig. 1(f)], the emergence of an additional hierarchy of even smaller minima at $T_G$ [inset of Fig. 1(f)] is generating a new relaxation process, which leads to an additional dielectric contribution [Fig. 1(d)]. In any case, the Gardner transition can be expected to leave a trace in the low-temperature behavior of the $\beta$ relaxation, fully consistent with our experimental findings.

In summary, we have provided detailed high-precision dielectric data on the JG $\beta$ relaxation of two typical glass formers extending down to unprecedented low temperatures. We have detected an extreme broadening of the $\beta$ relaxation when approaching He temperatures which only partly can be explained by the expected $1/T$ divergence. Irrespective of the detailed mechanism [scenario (i) or (ii) in the preceding paragraph], it seems natural to ascribe the found additional broadening far below $T_g$ to a further roughening or fractionalization of the existing sub-basins and to the theoretically predicted increase of the complexity of the energy landscape arising at the Gardner transition.

We thank Catalin Gainaru and Alexei Sokolov for helpful discussions.

# Supplemental Material

# Johari-Goldstein relaxation far below $T_g$:
# Experimental evidence for the Gardner transition in structural glasses?

K. Geirhos, P. Lunkenheimer*, and A. Loidl

*Experimental Physics V, Center for Electronic Correlations and Magnetism, University of Augsburg, 86159 Augsburg, Germany*

*e-mail: peter.lunkenheimer@physik.uni-augsburg.de

## 1. Temperature-dependent data

Figure S1 shows the temperature dependence of the dielectric loss of sorbitol and xylitol down to He temperature. The well-known $\alpha$ and secondary $\beta$ relaxations of these materials are nicely revealed leading to peaks or shoulders that shift with frequency. The glass transition shows up by a change of slope at $T_g$ but below $T_g$ there is no clear additional anomaly that is independent of frequency as expected for a well-defined phase transition.

## 2. Dielectric spectra of xylitol

Figure S2 shows the dielectric constant $\varepsilon'$ (a) and loss $\varepsilon''$ (b) of xylitol as measured below $T_g$ between 5 and 220 K. The inset shows a magnified view of the loss spectra at the lowest temperatures. The general behavior resembles that of sorbitol, which is discussed in detail in the main paper.

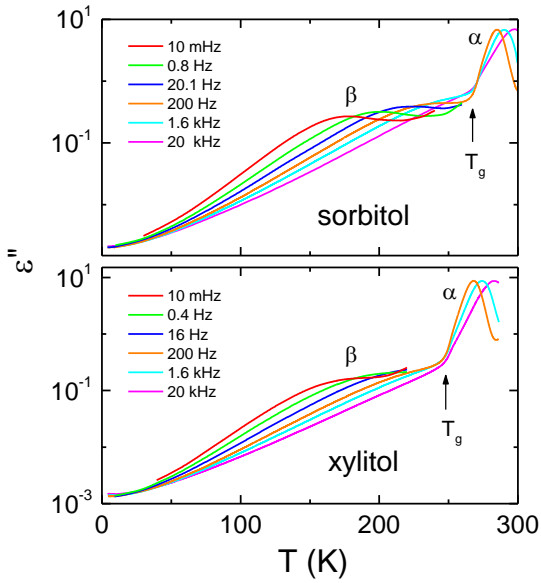

FIG. S1. Temperature-dependent dielectric loss of sorbitol (a) and xylitol (b) for different frequencies.

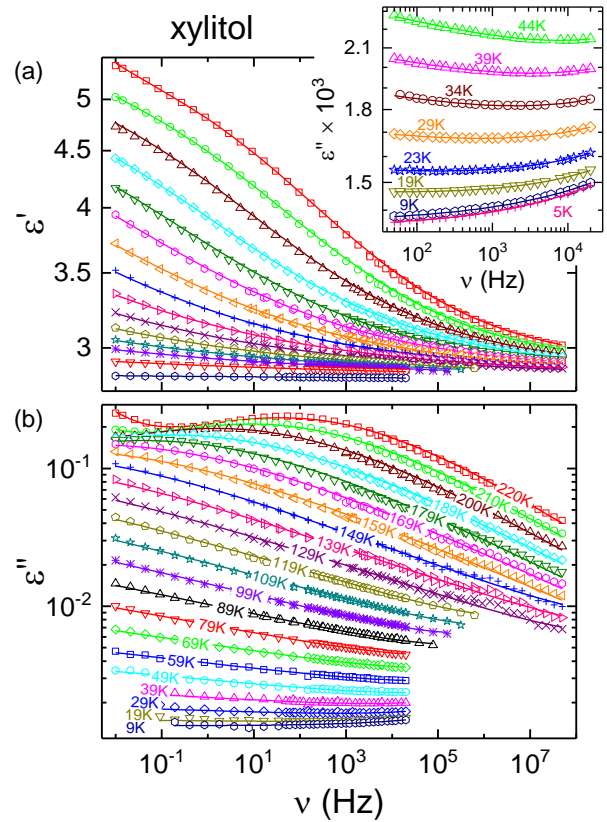

FIG. S2. Spectra of the dielectric constant (a) and loss (b) of sorbitol for various temperatures. The lines are fits by a combination of power laws and the CC function, analogous to the fits of the sorbitol spectra described in the main paper. For clarity reasons, in (a) at low temperatures curves are shown for less temperatures than in (b). The inset shows a magnified view of the loss in the region of the minimum.

## 3. Relaxation-time map

Figure S3 shows an Arrhenius plot of the temperature dependence of the $\alpha$- and $\beta$-relaxation times of sorbitol (the data at $T \geq 200$ K were already published in [1]). The solid lines are fits with a Vogel-Fulcher-Tammann law for the $\alpha$ relaxation [1] and with an Arrhenius law $\tau_\beta \propto \exp[E/(k_B T)]$ (with $E = 0.62$ eV the energy barrier) for the $\beta$ relaxation [1]. The slope of the fit line of the $\tau_\beta$ data is proportional to the energy barrier. The two dashed lines correspond to the lower and higher limiting energy barriers defining the half width of the distribution $g_E(E)$. The resulting width $W_\tau$ of the relaxation-time distribution $g_\tau(T)$ can be read off in this figure by the vertical distance of the two dashed lines. It becomes obvious, that $W_\tau$ becomes temperature dependent, despite a temperature-independent width of $g_E$, and that it strongly increases with decreasing temperature [2,3]. One should be aware that the lowest temperature covered by the abscissa in this figure is 118 K and that one would have to extend the abscissa to $1000/T = 200$ K$^{-1}$ in order to include the lowest temperature (5 K) investigated in this work. This explains the more than 100 decades wide $W_\tau$ values revealed in Fig. 3 at the lowest temperatures.

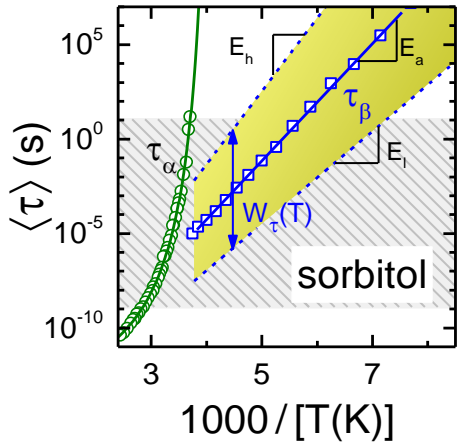

FIG. S3. (a) Temperature dependence of the $\alpha$- and $\beta$-relaxation times in sorbitol, shown in Arrhenius representation (symbols). The solid lines are fits with the Vogel-Fulcher-Tammann equation ($\alpha$ relaxation) or Arrhenius behavior ($\beta$ relaxation). The dashed lines indicate the Arrhenius behavior for the energy barriers $E_l$ and $E_h$ representing the lower and higher limiting barriers defining the half width of the distribution $g_E(E)$, respectively. The hatched area corresponds to the investigated frequency range using $\tau = 1/(2\pi\nu)$.